\documentclass[showpacs,preprintnumbers,10pt,twocolumn]{revtex4}
\usepackage{amssymb}
\usepackage{amsfonts}
\usepackage{amsmath}
\usepackage{graphicx}
\usepackage{dcolumn}
\usepackage{bm}
\usepackage{revsymb}

\input{tcilatex}

\begin{document}

\title{ Universal scaling laws of chaotic escape in dissipative multistable
systems subjected to autoresonant excitations}
\author{Ricardo Chac\'{o}n}
\affiliation{Departamento de F\'{\i}sica Aplicada, Escuela de Ingenier\'{\i}as
Industriales, Universidad de Extremadura, Apartado Postal 382, E-06071
Badajoz, Spain}
\date{\today}

\begin{abstract}
A theory concerning the emergence and control of chaotic escape from a
potential well by means of autoresonant excitations is presented in the
context of generic, dissipative, and multistable systems. Universal scaling
laws relating both the onset and lifetime of transient chaos with the
parameters of autoresonant excitations are derived theoretically using
vibrational mechanics, Melnikov analysis, and energy-based autoresonance
theory. Numerical experiments show that these scaling laws are robust
against both the presence of noise and driving re-shaping.
\end{abstract}

\pacs{05.45.Gg, 33.80.Gj}
\maketitle

\textit{Introduction.}$-$Escape from a potential well is an old problem with
wide-ranging implications where the interplay of noise, dissipation,
deterministic driving, and quantum uncertainty has given rise to unexpected
and intriguing phenomena such as coherent destruction of tunneling [1] and
stochastic resonance [2]. Other specific examples are pulse-shape-controlled
tunneling [3], shot-noise-driven escape in Josephson junctions [4], and
thermally induced escape [5]. While most previous investigations on driven
escape have restricted themselves to purely periodic drivings, chirped
excitations also have a demonstrated effectiveness. Chirped lasers, for
example, can reduce the intensity required for infrared multiphoton
dissociation of diatomic molecules to an experimentally realizable intensity
range [6]. And chirped optical pulses have been shown to enhance charge flow
in molecular-tunneling junctions [7]. In spite of the importance of chirped
excitations [8], to the best of the author's knowledge, laws governing (some
of) the associated escape scenarios remain to be revealed. The key to
understanding the aforementioned effectiveness of chirped excitations is
essentially the autoresonance (AR) mechanism: AR induced energy
amplification in nonlinear, driven, and deterministic systems occurs when
the system continuously adjusts its amplitude so that its instantaneous
nonlinear period matches the instantaneous driving period of the chirped
excitation. Initially studied in the context of a Hamiltonian description,
AR phenomena have been well known for about half a century and have been
observed in particle accelerators, planetary dynamics, atomic and molecular
physics, and nonlinear oscillators [9], to cite a few examples. Regarding
dissipative systems, an energy-based AR (EBAR) theory has recently been
proposed and applied to the case where the system crosses a separatrix
associated with its underlying integrable counterpart [10]. Since in such an
escape situation the appearance of transient chaos associated with the
occurrence of homoclinic bifurcations is an ubiquitous phenomenon, the
question naturally arises: How does AR control the chaotic escape scenario,
i.e., the onset and lifetime of transient chaos in generic dissipative
multistable systems?

\textit{Theory.}$-$In this Letter, this fundamental problem is studied in
the context of the family of systems%
\begin{equation}
\overset{..}{x}+dU/dx=-\delta \overset{.}{x}+\gamma \sin \left[ \Omega
\left( t\right) t\right] ,  \tag{1}
\end{equation}%
where $U(x)$ is a generic multistable potential (see Fig.~1), and $\Omega
\left( t\right) \equiv \omega +\alpha _{n}t^{n},n=1,2,...,$ is a
time-dependent frequency with $\alpha _{n}$ being the n-th-order sweep rate.
One expects [10] $\gamma \sin \left[ \Omega \left( t\right) t\right] $ to
behave as an effective autoresonant excitation whenever the sweep rate is
sufficiently low (adiabatic regime), which is the case considered throughout
the present paper in order to apply Melnikov analysis (MA) to autoresonant
excitations [11]. Also, the damping and autoresonant excitation terms are
taken to be small amplitude perturbations of the underlying integrable
system so as to deduce analytical expressions for the scaling laws relating
both the onset time $t_{i}$ and lifetime $\tau $ of transient chaos with the
parameters of the AR excitation from the MA [12] results. The application of
MA to any homoclinic (or heteroclinic) orbit $\left[ x_{h,j}\left( t\right) ,%
\overset{.}{x}_{h,j}\left( t\right) \right] $ of the unperturbed $\left(
\delta =\gamma =0\right) $ counterpart of Eq.~(1) involves calculating the
corresponding Melnikov function (MF)%
\begin{equation}
M_{j}\left( t_{0}\right) =-D+\gamma R_{0}\left( \omega ,t_{0}\right) +\gamma
\alpha _{n}R_{1}\left( \omega ,t_{0}\right) +O\left( \gamma \alpha
_{n}^{2}\right) ,  \tag{2}
\end{equation}%
where $D\equiv \delta \int_{-\infty }^{\infty }\overset{.}{x}%
_{h,j}^{2}\left( t\right) dt>0$, $R_{0}\left( \omega ,t_{0}\right) \equiv
\int_{-\infty }^{\infty }\overset{.}{x}_{h,j}\left( t\right) \sin \left[
\omega \left( t+t_{0}\right) \right] dt$, $R_{1}\left( \omega ,t_{0}\right)
\equiv \int_{-\infty }^{\infty }\left( t+t_{0}\right) ^{n}\overset{.}{x}%
_{h,j}\left( t\right) \cos \left[ \omega \left( t+t_{0}\right) \right] dt$.
For a purely periodic excitation $\left( \alpha _{n}=0\right) $, it is
assumed in the following that the system does not exhibit transient chaos
for a given set of parameter values, i.e., the MF (2) has no simple zeros: 
\begin{equation}
D>\gamma \left\vert \max_{t_{0}\in \ 
\mathbb{R}
}R_{0}\left( \omega ,t_{0}\right) \right\vert \equiv \gamma A_{0}\left(
\omega \right) ,  \tag{3}
\end{equation}%
where $A_{0}\left( \omega \right) $ represents a chaotic-threshold function
whose generic behaviour is shown in Fig.~1. \FRAME{ftbphFU}{3.1399in}{%
2.1314in}{0pt}{\Qcb{ (color online) Generic multistable potential $U(x)$ vs $%
x$, and energy levels corresponding to different separatrices (dashed
lines). The inset shows a generic chaotic threshold function $A_{0}\left( 
\protect\omega \right) $ vs $\protect\omega $ and the range $\left[ \protect%
\omega _{th,1},\protect\omega _{th,2}\right] $ in which homoclinic chaos is
expected [see Eq.~(3)].}}{}{figure1.eps}{\special{language "Scientific
Word";type "GRAPHIC";maintain-aspect-ratio TRUE;display "USEDEF";valid_file
"F";width 3.1399in;height 2.1314in;depth 0pt;original-width
4.0845in;original-height 2.8818in;cropleft "0.0475";croptop
"0.9282";cropright "0.9465";cropbottom "0.0673";filename
'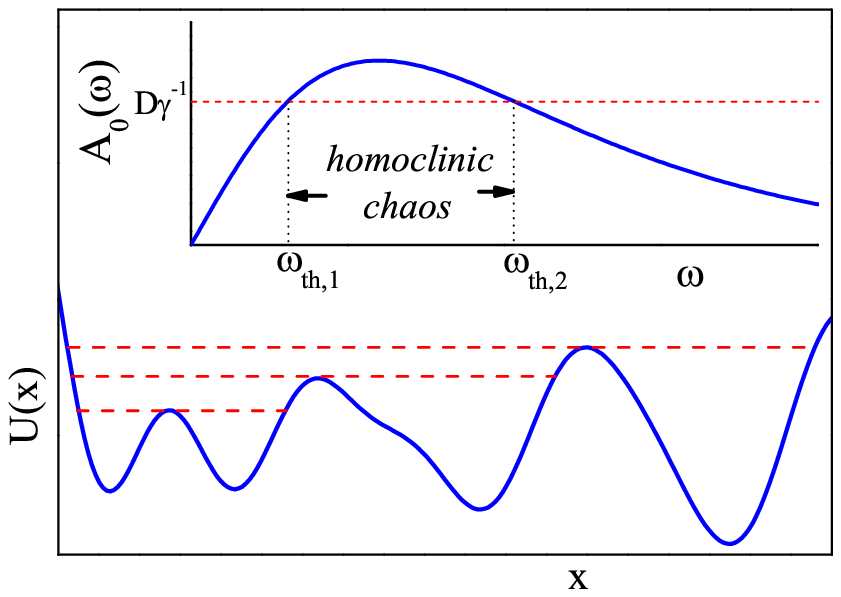';file-properties "XNPEU";}}Indeed, homoclinic chaos is not
possible for sufficiently small $\omega $ since a purely harmonic excitation
becomes a constant when $\omega \rightarrow 0$. For sufficiently high $%
\omega $, the dynamics can be analyzed using the vibrational mechanics
approach [13] by separating $x(t)=z(t)+\psi \left( t\right) $, where $z(t)$
represents the slow dynamics while $\psi \left( t\right) $ is the fast
oscillating term: $\psi \left( t\right) =\psi _{0}\cos \left( \omega
t+\varphi _{0}\right) $ with $\psi _{0}=\gamma \sqrt{\delta ^{2}+\omega ^{2}}%
/\left( \omega \delta ^{2}-\omega ^{3}\right) $, $\varphi _{0}=\arctan
\left( \delta /\omega \right) $. On averaging out $\psi \left( t\right) $
over time, the slow reduced dynamics of the system becomes 
\begin{eqnarray}
\overset{..}{z}+dV/dz &=&-\delta \overset{.}{z},  \notag \\
dV/dz &\equiv &T^{-1}\int_{0}^{T}g\left[ z+\psi _{0}\cos \left( \omega
t+\varphi _{0}\right) \right] dt,  \TCItag{4}
\end{eqnarray}%
where $g\left( x\right) \equiv dU(x)/dx,T\equiv 2\pi /\omega $, i.e., that
of a \textit{purely} damped system, and hence homoclinic chaos is not
possible when $\omega \rightarrow \infty $. Note that this demonstrates that
equilibria are the \textit{only} attractors of the system for $\alpha _{n}>0$%
. Thus, one concludes that the properties $A_{0}\left( \omega \rightarrow
0,\infty \right) =0,A_{0}\left( \omega \right) \geqslant 0$ imply via the
extreme value theorem (Weierstrass' theorem [14]) that the generic
chaotic-threshold function $A_{0}\left( \omega \right) $ presents at least
one maximum (the case shown in the inset of Fig.~1). Equation (2) indicates
that the MF $M_{j}\left( t_{0}\right) $ has simple zeros (at sufficiently
large values of $t_{0}$) for any positive value of the sweep rate because of
the factor $\left( t+t_{0}\right) ^{n}$ appearing in the definition of $%
R_{1}\left( \omega ,t_{0}\right) $. Physically, this means that after a
sufficiently long time, $t_{i}$, which depends upon the system's parameters
and initial conditions, the instantaneous frequency of the autoresonant
excitation reaches a threshold value $\Omega \left( t_{i}\right) =\omega
_{th,1}$ that is the lowest frequency satisfying the relationship $D/\gamma
=A_{0}\left( \omega _{th,1}\right) $ (see Fig.~1, inset), i.e., the
threshold condition for the onset of chaotic behaviour. It is worth noting
that this occurs for any initial conditions because of the AR-induced
increase of the system's energy. Thus, the condition $\Omega \left(
t_{i}\right) =\omega _{th,1}$ implies the scaling law%
\begin{equation}
t_{i}\sim \left( \omega _{th,1}-\omega \right) ^{1/n}\alpha _{n}^{-1/n}, 
\tag{5}
\end{equation}%
for the onset time of transient chaos. Similarly, the lifetime $\tau $ of
the chaotic transients can be estimated from the instantaneous frequency $%
\Omega \left( t_{i}+\tau \right) =\omega _{th,2}$ that is the lowest
frequency satisfying the relationships $D/\gamma =A_{0}\left( \omega
_{th,2}\right) ,\omega _{th,2}>\omega _{th,1}$ (see Fig.~1, inset), i.e.,%
\begin{equation}
\tau \sim \left[ \left( \omega _{th,2}-\omega \right) ^{1/n}-\left( \omega
_{th,1}-\omega \right) ^{1/n}\right] \alpha _{n}^{-1/n}.  \tag{6}
\end{equation}%
We see that the universal scaling laws (5) and (6) are \textit{inverse-power}
laws containing the\textit{\ }two parameters $\left( \omega ,\alpha
_{n}\right) $ that control the autoresonant excitation while the critical
exponent is the inverse of the chirp order $n$. Notably, these scaling laws
also contain the dependence upon the particular potential, initial potential
well, and dissipation and excitation strengths through the threshold
frequencies $\omega _{th,1},\omega _{th,2}$. Since the mid 1980's, critical
exponents of chaotic transients have been discussed theoretically in the
context of crises in dissipative maps [15]. The present theory establishes
that an inverse-power law remains valid also for dissipative multistable
flows subjected to autoresonant excitations. The connection between such
results for maps and the present ones can be readily understood by assuming
that for a purely periodic excitation $\left( \alpha _{n}=0\right) $ there
exists a chaotic attractor for a certain frequency $\left( \omega
_{th,1}<\omega <\omega _{th,2}\right) $ instead of a periodic attractor $%
\left( \omega _{th,1}>\omega \right) $ (see Fig.~1, inset). According to the
above discussion, the chaotic attractor disappears for $\alpha _{n}>0$ via a
boundary crisis and chaotic transients appear instead, with $\alpha =\alpha
_{n,c}\equiv 0$ being the critical value for the crisis. Therefore, since
the non-existence of simple zeros of the MF is a sufficient condition for
the disappearance of transient chaos, the lifetime of these chaotic
transients follows the universal inverse-power law%
\begin{equation}
\tau \sim \left( \omega _{th,2}-\omega \right) ^{1/n}\alpha _{n}^{-1/n}. 
\tag{7}
\end{equation}%
Now, applying EBAR theory one can deduce scaling laws relating both the
onset time and lifetime with the autoresonant excitation amplitude. In
particular, for Duffing-like potentials one has that the optimal amplitude
scales as $\gamma \sim \left[ 3^{n}\left( n+1\right) !\right] ^{3/\left(
2n+2\right) }\alpha _{n}^{3/\left( 2n+2\right) }$ [10] and hence the scaling
law (7) for example becomes $\tau \sim \left[ \left( \omega _{th,2}-\omega
\right) \left( n+1\right) !\right] ^{1/n}\gamma ^{-(2n+2)/(3n)}$, where one
sees that the critical exponent ranges from 4/3 for a linear chirp to 2/3
for the highest-order chirps.

\textit{Robustness vs noise and re-shaping.}$-$Numerical experiments
confirmed the accuracy of the above scaling laws in different systems.
Representative results corresponding to a dimensionless Duffing oscillator $%
\overset{..}{x}=x-x^{3}-\delta \overset{.}{x}+\gamma \limfunc{sn}\left[
2K(m)\Omega \left( t\right) t/\pi ;m\right] +\sigma N\left( 0,1\right) $ are
shown in Figs.~2--5 for illustrative purposes. Here, $\limfunc{sn}\left(
\cdot ;m\right) $ is the Jacobian elliptic function of parameter $m\in \left[
0,1\right] $ with $K(m)$ being the complete elliptic integral of the first
kind. One has $\limfunc{sn}\left( \cdot ;m=0\right) =\sin \left( \cdot
\right) $ while, in the other limit, $\limfunc{sn}\left( \cdot ;m=1\right) $
is the square-wave function [16]. Also, $\sigma N\left( 0,1\right) $ is a
white-noise term with $N\left( 0,1\right) $ being a random variable with
Gaussian distribution, zero mean, and variance 1, while $\sigma >0$ controls
the noise strength. For the purely deterministic $\left( \sigma =0\right) $
case of a sinusoidal $\left( m=0\right) $ excitation with a linear chirp $%
\left( n=1\right) $, for example, and after calculating the resulting
integrals by residues, one obtains the corresponding MF: $M_{Duffing}^{\pm
}\left( t_{0}\right) =-D\mp B_{0}\cos \left( \omega t_{0}\right) \mp \alpha
_{1}\left[ B_{1}t_{0}\cos \left( \omega t_{0}\right) -\left(
B_{2}+B_{0}t_{0}^{2}\right) \sin \left( \omega t_{0}\right) \right] +O\left(
\gamma \alpha _{1}^{2}\right) $, where the sign $+\left( -\right) $
corresponds to the right (left) homoclinic orbit of the unperturbed Duffing
oscillator $\left( \delta =\gamma =0\right) $, and $D\equiv 4\delta /3$, $%
B_{0}\equiv \sqrt{2}\pi \gamma \omega \func{sech}\left( \pi \omega /2\right) 
$, $B_{1}\equiv 4\sqrt{2}\gamma \left[ \left( \pi /2\right) -\left( \pi
^{2}\omega /4\right) \tanh \left( \pi \omega /2\right) \right] \func{sech}%
\left( \pi \omega /2\right) $, $B_{2}\equiv \sqrt{2}\gamma \left( \pi
^{2}/8\right) \func{sech}^{3}\left( \pi \omega /2\right) \left[ 4\sinh
\left( \pi \omega \right) -\pi \omega \cosh \left( \pi \omega \right) +3\pi
\omega \right] $. \FRAME{ftbphFU}{3.1914in}{2.5474in}{0pt}{\Qcb{(color
online) Onset time (circles), $t_{i}$, and lifetime (squares, inset), $%
\protect\tau $, of transient chaos corresponding to a dimensionless Duffing
oscillator (see the text) for $m=0,\protect\sigma =0,\protect\delta =0.5,%
\protect\gamma =0.4$, and $\protect\omega =0.493$. Hence, $\protect\omega %
_{th,2}=1.11736,\protect\omega _{th,1}=0.49371$ from $M_{Duffing}^{\pm
}\left( t_{0}\right) $ (see the text). Solid lines indicate fits according
to the scaling laws (5) and (6), respectively.}}{}{figure2.eps}{\special%
{language "Scientific Word";type "GRAPHIC";maintain-aspect-ratio
TRUE;display "USEDEF";valid_file "F";width 3.1914in;height 2.5474in;depth
0pt;original-width 4.285in;original-height 3.5116in;cropleft
"0.0439";croptop "0.9317";cropright "0.9471";cropbottom "0.0537";filename
'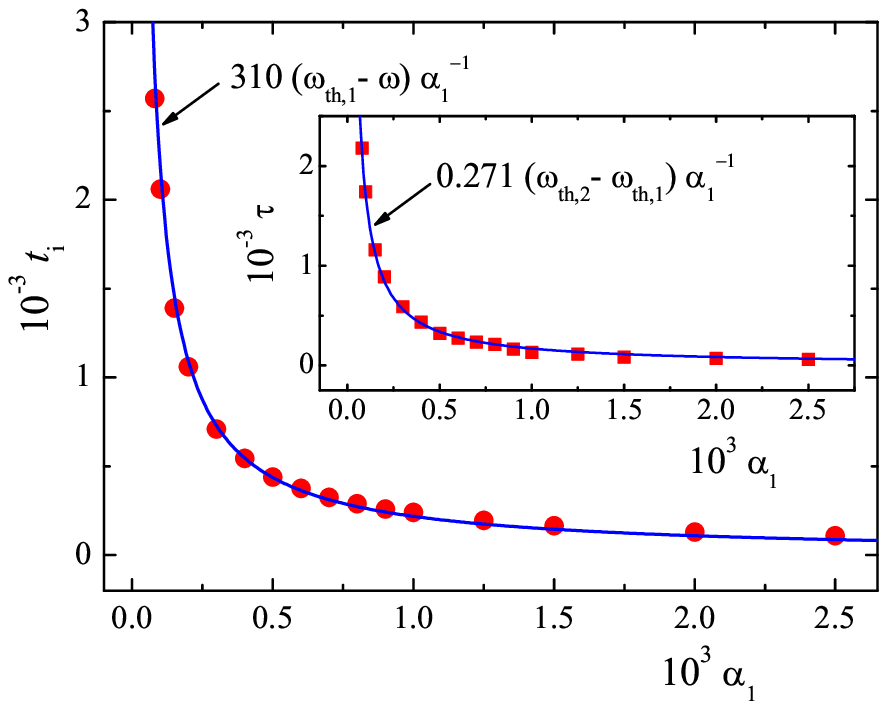';file-properties "XNPEU";}}The results for the case of a
periodic (chaotic) attractor existing at $\alpha _{n}=0$ are shown in Fig.~2
(Fig.~4). \FRAME{ftbphFU}{3.5807in}{2.7115in}{0pt}{\Qcb{(color online)
Autoresonant response of a dimensionless Duffing oscillator for a linear
chirp $\protect\alpha _{1}=3\times 10^{-4}$ (see the text). (a) Position vs
time and final equilibrium (dashed line). (b) Energy and average energy
(over a few periods $2\protect\pi /\protect\omega $, thick black line) vs
time. (c) Phase space trajectory and period-1 attractor existing at $\protect%
\alpha _{n}=0$ (thick black line). (d) Energy vs position (dashed line
indicates the separatrix energy level as in version (b)). Remaining
parameters are the same as in Fig.~2.}}{}{figure3.eps}{\special{language
"Scientific Word";type "GRAPHIC";maintain-aspect-ratio TRUE;display
"USEDEF";valid_file "F";width 3.5807in;height 2.7115in;depth
0pt;original-width 4.0411in;original-height 3.1541in;cropleft "0";croptop
"1";cropright "1.0332";cropbottom "0";filename
'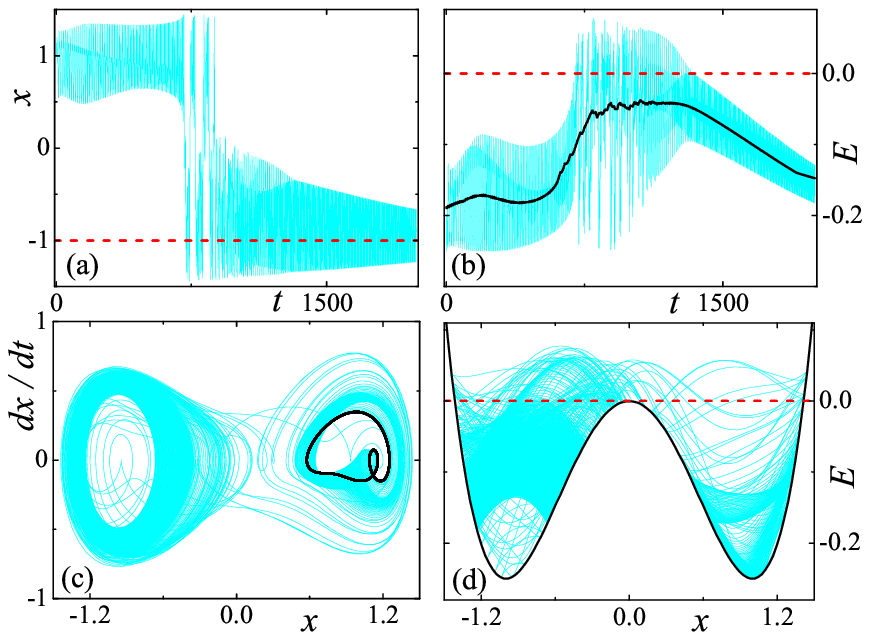';file-properties "XNPEU";}}While a linear chirp is
considered in Figs.~2 and 3, Fig.~4 shows the results for quadratic and
cubic chirps. \FRAME{ftbphFU}{3.7511in}{2.8641in}{0pt}{\Qcb{(color online)
Lifetime of transient chaos corresponding to a dimensionless Duffing
oscillator (see the text) for quadratic ($n=2$, squares) and cubic ($n=3$,
circles) chirps. Solid lines indicate fits according to the scaling law (7).
Also shown is the chaotic attractor existing at $\protect\alpha _{2,3}=0$.
Fixed parameters: $m=0,\protect\sigma =0,\protect\delta =0.154,\protect%
\gamma =0.2,\protect\omega =1.1$. Hence, $\protect\omega _{th,2}=1.71404$
from $M_{Duffing}^{\pm }\left( t_{0}\right) $ (see the text).}}{}{figure4.eps%
}{\special{language "Scientific Word";type "GRAPHIC";maintain-aspect-ratio
TRUE;display "USEDEF";valid_file "F";width 3.7511in;height 2.8641in;depth
0pt;original-width 4.3693in;original-height 3.4681in;cropleft "0";croptop
"1";cropright "1.0420";cropbottom "0";filename
'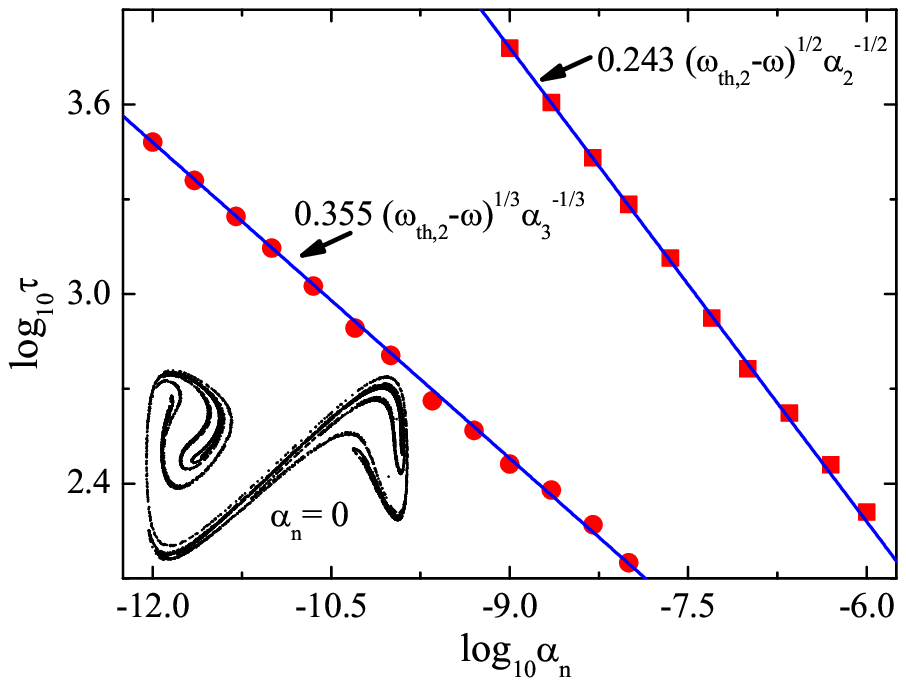';file-properties "XNPEU";}}Figures 2 and 4 show excellent
agreement between numerical results and the predicted scaling laws for both
the onset time and the lifetime of transient chaos. Figure 3(b) shows the
AR-induced increase of energy over time until the separatrix energy level is
reached, indicating the onset of transient chaos (see Figs.~3(a) and 3(d)),
followed by a decrease of energy when the dynamics becomes effectively
purely dissipative (see Fig.~3(c)). Note that the average energy remains
roughly constant during transient chaos. The robustness of the scaling laws
versus both the presence of additive noise and re-shaping of the
autoresonant excitation is shown in Fig.~5. Note that, for the elliptic
excitation, $\omega _{th,i},i=1,2$, are functions of $m$, which is the
reason for the different prefactors in the fits for $m=0.995$ and $%
m=1-10^{-14}$ (Fig.~5, bottom).\FRAME{ftbphFU}{3.1736in}{2.5128in}{0pt}{\Qcb{%
(color online) Lifetime of transient chaos corresponding to a dimensionless
Duffing oscillator (see the text) in the presence of noise (top panel, $m=0,%
\protect\gamma =0.4$) and subjected to elliptic excitations (bottom panel, $%
\protect\sigma =0,\protect\gamma =0.3$). Solid lines indicate fits according
to the scaling law (6). Fixed parameters: $\protect\delta =0.5,\protect%
\omega =0.493$.}}{}{figure5.eps}{\special{language "Scientific Word";type
"GRAPHIC";maintain-aspect-ratio TRUE;display "USEDEF";valid_file "F";width
3.1736in;height 2.5128in;depth 0pt;original-width 4.6265in;original-height
3.6526in;cropleft "0";croptop "1";cropright "1";cropbottom "0";filename
'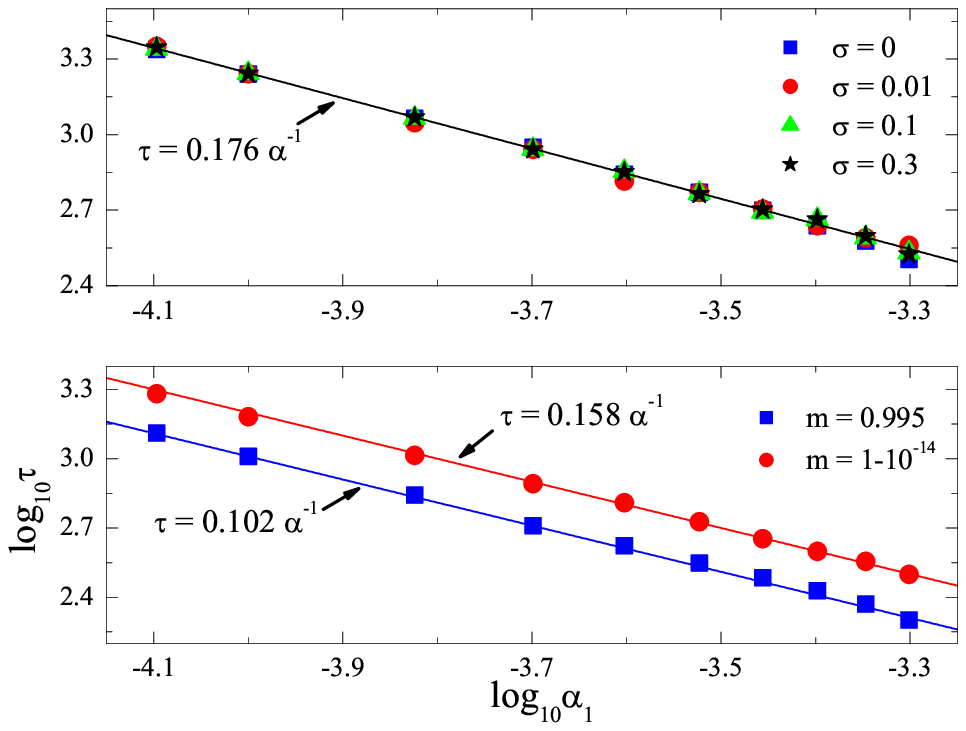';file-properties "XNPEU";}}

\textit{Conclusions.}$-$In sum, universal inverse-power laws relating both
the onset and lifetime of transient chaos and the parameters of
escape-inducing autoresonant excitations have been theoretically derived for
generic, dissipative, and multistable flows. Numerical simulations showed
the robustness of these scaling laws against both the presence of additive
noise and driving re-shaping. Since the critical exponents were found to
solely depend on the chirp's order, such scaling laws are expected to remain
valid for even more general dissipative systems including spatiotemporal
chaotic systems [17]. The present results can be readily tested
experimentally (for example in mechanical and laser systems), and can find
application to optimally control elementary dynamic processes characterized
by chaotic escapes from a potential well, such as diverse atomic and
molecular processes, transport phenomena in dissipative lattices, and
control of the electron dynamics in quantum solid-state devices.

The author thanks Francisco Balibrea for stimulating discussions. Work
partially supported by the Spanish MCyT through project FIS2008-01383.


\begin{thebibliography}{99}
\bibitem{16} Note that, with the instantaneous period constant, solely the
excitation shape is varied by increasing $m$ from $0$ to $1$, and there is
thus a smooth transition from a sine function to a square wave.
\end{thebibliography}
\end{document}